\documentclass[12pt,twoside]{article}   
\usepackage[super,sort,comma]{natbib}

\usepackage{fancyhdr}		

\usepackage[section]{placeins}   %

\usepackage{graphicx}

\makeatletter \renewcommand\@biblabel[1]{$^{#1}$} \makeatother
\newcommand{\cen}[1]{\begin{center} #1 \end{center}}

\usepackage[mathlines]{lineno}

\usepackage{hyperref}
\hypersetup{ colorlinks,
    citecolor=blue,
    filecolor=blue,
    linkcolor=blue,
    urlcolor=blue
}

\usepackage{xcolor}

\definecolor{gray}{rgb}{0.6,0.6,0.6}
\definecolor{red}{rgb}{0.85,0,0}
\definecolor{green}{rgb}{0,0.85,0}
\definecolor{blue}{rgb}{0,0,0.85}
\definecolor{beige}{rgb}{0.92,0.87,0.78}
\usepackage[all]{hypcap}    

\begin{document}

\cen{\sf {\Large {\bfseries The innovative $^{52g}$Mn for PET imaging: production cross section modeling and dosimetric evaluation } \\  
\vspace*{10mm}
F. Barbaro$^{1,2,*}$, L. Canton$^1$, M.P. Carante$^{2,3}$, A. Colombi$^{2,3}$, \\ L. De Nardo$^{1,4,*}$, A. Fontana$^3$, L. Meléndez-Alafort$^{5}$} \\
$^1$ INFN, Sezione di Padova, Padova, Italy, $^2$ Dipartimento di Fisica dell'Universit\`a di Pavia, Pavia, Italy, $^3$ INFN, Sezione di Pavia, Pavia, Italy,
$^4$ Dipartimento di Fisica e Astronomia dell'Universit\`a di Padova, Padova, Italy,
$^5$ Istituto Oncologico Veneto IOV IRCCS, Padova, Italy
}

\pagenumbering{roman}
\setcounter{page}{1}
\pagestyle{plain}
*Corresponding authors: francesca.barbaro@pd.infn.it;  laura.denardo@unipd.it








\definecolor{ForestGreen}{RGB}{34,139,34}
\newcommand{\andrea}[1]{\textcolor{ForestGreen}{#1}}
\newcommand{\luciano}[1]{\textcolor{red}{#1}}
\newcommand{\francesca}[1]{\textcolor{magenta}{#1}}
\newcommand{\mario}[1]{\textcolor{blue}{#1}}
\newcommand{\alessandro}[1]{\textcolor{orange}{#1}}
\newcommand{\laurad}[1]{\textcolor{brown}{#1}}
\newcommand{\lauram}[1]{\textcolor{cyan}{#1}}

\begin{abstract}
\noindent {\bf Background:} Manganese is a paramagnetic element suitable for magnetic resonance imaging (MRI) of neuronal function, however high concentrations of Mn$^{2+}$ can cause neurological disorders. $^{52g}$Mn appears a valid alternative as PET (positron emission tomography) imaging agent, to obtain information similar to that delivered by MRI but using trace levels of Mn$^{2+}$, thus reducing its toxicity. Recently, the reaction $^{nat}$V($\alpha$,x)$^{52g}$Mn has been proposed as a possible alternative to the standard $^{nat}$Cr($p$,x)$^{52g}$Mn one, but improvements in the modeling were needed to better compare the two production routes. \\
{\bf Purpose:} 
This work focuses on the development of precise simulations and models to compare the $^{52g}$Mn production from both reactions in terms of amount of activity and radionuclidic purity (RNP), as well as in terms of dose increase (DI) due to the co-produced radioactive contaminants, with respect to a pure $^{52g}$MnCl$_2$ injection. \\
{\bf Methods:}
The nuclear code Talys has been employed to optimize the $^{nat}$V($\alpha$,x)$^{52g}$Mn cross section by tuning the parameters of the microscopic level densities. Thick target yields have been calculated from the expression of the rates as energy convolution of cross sections and stopping powers, and finally integrating the time evolution of the relevant decay chains. Dosimetric assessments of [$^{xx}$Mn]Cl$_2$ have been accomplished with OLINDA software 2.2.0 using female and male phantoms. At the end, the yield of $^{xx}$Mn radioisotopes estimated for the two production routes have been combined with the dosimetric results, to assess the DI at different times after the end of the irradiation. \\
{\bf Results:} Good agreement was obtained between cross sections calculations and measurements. The comparison of the two reaction channels suggests that $^{nat}$V($\alpha$,x)$^{52g}$Mn leads to higher yield and higher purity, resulting in a less harmful impact on patients’ health in terms of DI. \\
{\bf Conclusions:} 
Both $^{nat}$V($\alpha$,x) and $^{nat}$Cr($p$,x) production routes provide clinically acceptable $^{52g}$MnCl$_2$ for PET imaging. However, the $^{nat}$V($\alpha$,x)$^{52g}$Mn reaction provides a DI systematically lower than the one obtainable with $^{nat}$Cr($p$,x)$^{52g}$Mn and a longer time window in which it can be used clinically (RNP $\ge$ 99\%). \\



\end{abstract}

\section{Introduction}
Manganese is an essential element for all beings because it participates as a cofactor in numerous enzymatic processes. The most stable oxidation state is manganese (II), the divalent manganese cation (Mn$^{2+}$). It behaves similarly to calcium and can be diffused through voltage-gated calcium channels in the brain, heart and pancreas. Increased activity in these tissues leads to an increased influx of Mn$^{2+}$; therefore, this cation can be used to monitor cell activity \cite{saar}. In addition, manganese is a unique tool for obtaining in vivo images of neural pathways because Mn$^{2+}$ possesses strong paramagnetic properties and can be transported along the axons in an anterograde manner and cross synapses \cite{napieczynska}. Therefore, Mn$^{2+}$ has been used to obtain manganese-enhanced magnetic resonance imaging (MEMRI). This imaging technique enables early detection of neuronal function, intracellular ion balance, and axonal transport; nevertheless, it  has not been widely used in clinical practice since the high doses of Mn$^{2+}$ have been found to lead to a neurological disorder called manganism, with psychiatric and neurological syndromes similar to those of the Parkinson’s disease \cite{yang}. Positron emission tomography (PET) offers superior contrast sensitivity compared to MRI and allows all typical information from MEMRI to be obtained using trace levels of Mn$^{2+}$ thus minimizing its toxic effects \cite{graves}.
Several radioisotopes of manganese have been studied as radiotracers for PET imaging \cite{coenen}. Among them $^{52g}$Mn appears the most suitable for this scope because it decays with a low maximum positron energy of about 0.6 MeV, that makes PET imaging resolution similar to that of $^{18}$F \cite{saar}. In addition, $^{52g}$Mn half-life ($t_{1/2}$ = 5.6 day) is long enough for in vivo imaging of cell tracking at time points as long as days or even weeks post-injection \cite{wooten}. 

The  best known production route is based on the reaction $^{nat}$Cr($p$,x)$^{52g}$Mn.
El Sayed et al \cite{sayed} recently measured  the cross-sections of $^{52g}$Mn, and the co-produced contaminants $^{54}$Mn and $^{51}$Cr with the goal of optimizing the routine production of $^{52g}$Mn. Experimental data were in good agreement with calculations obtained using the standard simulations with TALYS code \cite{talys}, and with previously published data. In addition cross-section data were useful to improve the available nuclear data sets for each radionuclide involved.
Recently, our group proposed an alternative reaction $^{nat}$V($\alpha$,x)$^{52g}$Mn with potentially higher yield and better radionuclidic purity \cite{colombi}. The study involved also considerations on the use of enriched targets which perform more efficiently but require more expensive materials and complicated target-recovery approaches. The study revealed that the experimental data sets for the $^{nat}$V($\alpha$,x)$^{52g}$Mn reaction are very scattered and need to be improved, and that the standard simulations with nuclear reaction codes, such as TALYS, FLUKA and EMPIRE, are not fully adequate. In this work the outcomes of FLUKA and EMPIRE codes are not taken into account since they do not provide an easy way to vary the parameters of their models. Instead, the study exploited the specific versatility of TALYS, which incorporates six different level-density models (three phenomenological and three based on microscopic theories) and four different pre-equilibrium approaches (three based on the exciton model \cite{cline} and one on the fully quantum-mechanical multi-step approach \cite{feshbach}). The multiple combination suggested a statistical treatment of the model variability, that leads to consider an inter-quartile band covering the area between the  25\% and 75\% percentiles of the entire set of possible calculations. This band gives and estimation of the spread (or theoretical indetermination, or "error") one gets for selecting one model instead of another one of the whole set. The median line of this inter-quartile band was chosen as the most representative line of the statistical assortment of calculations and the half width of the band was taken as an indication of the expected theoretical error. However, this approach to the $^{nat}$V($\alpha$,x)$^{52g}$Mn production cross-section \cite{colombi} was not optimal, since all the considered TALYS calculations had the tendency to shift the peak at lower energy, and slightly overestimate its maximum value. On the other hand, the spread of the experimental data appears too large for attempting a reliable evaluation of the cross section using interpolation techniques. In this work, we found a way to optimize the description of this cross-section using the microscopic models incorporated into the TALYS code, and by tuning the nuclear level density parameters of the $^{52}$Mn compound. The procedure follows an approach developed recently for the optimization of the $^{nat}$V(p,x)$^{47}$Sc cross section \cite{barbaro} and described in \ref{sect:x-sect-Mns}.

In view of possible applications of manganese radionuclides in medical imaging, it is important to assess the impact of the produced radionuclides in terms of dose released to the patient, including the contribution of the radioisotopic contaminants. The fraction of co-produced contaminants characterizes
the production route and represents a significant indicator to compare different approaches. De Nardo et al. analyzed the effective dose burden due to the use of $^{52g}$Mn as brain tracer under the simple MnCl$_2$ chemical compound with computational dosimetry codes \cite{denardo2019}. It was found that the radiation dose released by $^{52g}$Mn may be quite significant due to its relatively long physical half-life and the emission of high-energy $\gamma$ rays. The main focus of that publication was a comparison of the dosimetric properties of $^{52g}$Mn and $^{51}$Mn, a shorter half-life positron emitter ($t_{1/2}$ = 45.59 min, $\beta^{+}$=97.1\%, E($\beta^{+}$) avg = 970.2 keV), also suitable for PET imaging, despite its lower spatial resolution characteristics compared to $^{52g}$Mn. In this work, the main interest was the assessment of the quality of the produced $^{52g}$Mn, not only in terms of radionuclidic purity, but also by considering the dose increase due to the presence of radioisotopic contaminants, which remain in the final product because chemical purification is not possible. For these reasons, dosimetric calculations have been extended to [$^{xx}$Mn]Cl$_2$ labelled with the Mn radioisotopic impurities expected to be co-produced by the nuclear reactions investigated for $^{52g}$Mn production.

\section{Methods}

\subsection{Cross-section and yield evaluations}
\label{sect:x-sect-Mns}
The TALYS code was used to improve the model reproduction of the dataset, following the approach discussed previously by our group \cite{barbaro} and herein described in details for the cross section $^{nat}$V($\alpha$,x)$^{52g}$Mn. To this aim, the level density model based on microscopic Hartree-Fock-Bogoliubov theory with the Gogny force, incorporating a temperature dependence in the level density \cite{hilaire2012}, has been considered. This model is denoted as \emph{ldmodel 6} and is the latest level density model implemented in the TALYS code. Concerning the optical model, we have adopted the default options for the nucleon-nucleus optical model and for the $\alpha$-nucleus optical potential, corresponding to the model reported by Avrigeanu et al \cite{avrigeanu2014}. The pre-equilibrium reaction mechanisms were incorporated using the exciton model,  where the different exciton states were coupled with transition rates calculated numerically starting from the imaginary part of the optical potential. This references to \emph{preeqmode 3} in the TALYS code. With this selection of models, and following the approach reported previously \cite{barbaro}, the tabulated microscopic level density of the \emph{ldmodel 6} has been parameterized according to the following transformation
\begin{equation}
\rho(E,J,\pi) = exp(c \sqrt{E-p}) \rho_{HF}(E-p,J,\pi) \,  ,
\label{eq:1}
\end{equation}
where $\rho_{HF}(E,J,\pi)$ is the level density from the microscopic Hartree-Fock calculation.
The $c$ and $p$ parameters have to be optimized to obtain a cross-section in better agreement with the data and this has been performed by introducing a two-dimension search grid that minimizes the discrepancies between calculated and measured cross sections.
Once the cross section optimization has been performed, it is important also to check its impact on the level density by comparing the resulting level cumulatives with the ones derived experimentally \cite{ripl} (see  section \ref{result_xs}).

The optimized cross section can be employed to evaluate the number of radionuclides produced under specific irradiation conditions, and from there activities, yields, and purities. The computational approach considered has been extensively discussed by Canton et al \cite{canton-fontana} and outlined here for simplicity.

First, the most promising energy window for the production of $^{52g}$Mn at low contamination shall be identified. The rate $R$ of production of a radionuclide from a beam colliding on a target with a certain thickness can be derived from the expression

\begin{equation}
	R = \frac{I_0}{z_{proj}|e|}\frac{N_a}{A}\int_{E_{out}}^{E_{in}}\sigma(E)\left(\frac{dE}{\rho_tdx}\right)^{-1}dE \, ,
\end{equation}

where $I_0$ is the beam current, $z_{proj}$ is 2 for a completely ionized $^4He$ beam, $e$ the electron charge, $N_a$ the Avogadro number, $A$ the atomic mass of the target element, $E_{in}$ and $E_{out}$ the energy of the beam hitting the target and the one leaving the target after traveling through its thickness, respectively. $\sigma(E)$ is the production cross section for the nuclide, $\rho_t$ the target density and dE/dx the stopping power of the projectile in the target, calculated with the Bethe-Bloch formula \cite{leo}. 

Once the rate for all the Mn radionuclides of interest are calculated, the time evolution of the number of nuclei, and hence the activity of a specific radionuclide can be obtained, during and after the irradiation, by means of standard treatment of the decay evolution of each isotope through the Bateman equations. Each considered manganese radionuclide decays to different chemical elements, except $^{52m}$Mn, which decays in $^{52g}$Mn with a branching ratio of 1.75$\%$ and with a 21.1 minutes half-life. The main decay data of interest for the present analysis are reported in tables \ref{tab:decay_1} and \ref{tab:decay_2}. These data were extracted with the software package
DECDATA provided by the ICRP 107 publication \cite{ICRP107}.

\begin{table}
\begin{center}
\footnotesize
\begin {tabular}{|c|c|c|c|c|c|c|}
\hline
Radioisotope & \multicolumn{3}{|c|}{$^{51}$Mn; Half-life: 46.2 m; Decay: EC, $\beta^{+}$} & \multicolumn{3}{|c|}{$^{52g}$Mn; Half-life: 5.591 d; Decay: EC, $\beta^{+}$} \\ \hline
Radiations & Yield (/nt) & E (MeV/nt) & Mean E (MeV) & Yield (/nt) & E (MeV/nt) & Mean E (MeV) \\ \hline
Gamma rays & 5.242$\times 10^{-3}$ & 5.451 $\times 10^{-3}$ & 1.040 & 3.000 & 3.154 & 1.051 \\ \hline
X rays & 1.943$\times10^{-1}$ & 3.968$\times 10^{-5}$ & 2.042$\times 10^{-4}$  & 4.697 & 9.588$\times 10^{-4}$ & 2.042$\times 10^{-4}$ \\ \hline
Annh. Photons &	1.942 & 9.922$\times 10^{-1}$ & 5.11$\times 10^{-1}$ & 5.932$\times 10^{-1}$ & 3.031$\times 10 ^{-1}$ & 5.11$\times 10^{-1}$ \\ \hline
Tot photons	& - & 0.9977 & - & - & 3.4585 & - \\ \hline
$\beta^{+}$	& 9.709$\times 10 ^{-1}$ & 9.342$\times 10^{-1}$ & 9.623$\times 10^{-1}$ & 2.966$\times 10^{-1}$ & 7.172$\times 10^{-2}$ & 2.418$\times10^{-1}$  \\ \hline
$\beta^{-}$ & - & - & - & - & - & - \\ \hline
IC electrons & 1.130$\times 10^{-6}$ & 9.231$\times 10^{-7}$ & 8.165$\times 10^{-1}$ & 6.004$\times 10^{-4}$ & 5.109$\times 10^{-4}$ & 8.509$\times 10^{-1}$ \\ \hline
Auger electrons & 1.580$\times 10^{-1}$ & 1.167$\times 10^{-4}$ & 7.388$\times 10^{-4}$ & 3.818 & 2.820$\times 10^{-3}$ & 7.386$\times 10^{-4}$ \\ \hline
Tot electrons &	- & 0.9344 & - & - & 0.0750 & - \\ \hline
Tot	& - & 1.9321 & - & - & 3.5335 & - \\ \hline

\end{tabular}
\caption{Main decay data of the $^{51}$Mn and $^{52g}$Mn radionuclides extracted with the software package DECDATA provided by the ICRP 107 publication. For each radiation type, the yield or number per nuclear transformation (nt), the total emitted energy (MeV per nt), the mean (average) emitted energy (MeV) are reported.}
\label{tab:decay_1}
\end{center}
\end{table}

\begin{table}
\begin{center}
\footnotesize
\begin {tabular}{|c|c|c|c|c|c|c|}
\hline
Radioisotope & \multicolumn{3}{|c|}{$^{53}$Mn; Half-life: 3.7$\times$ 10$^{6}$ y; Decay: EC} & \multicolumn{3}{|c|}{$^{54}$Mn; Half-life: 312.12 d; Decay: EC, $\beta^{+}$, $\beta^{-}$} \\ \hline
Radiations & Yield (/nt) & E (MeV/nt) & Mean E (MeV) & Yield (/nt) & E (MeV/nt) & Mean E (MeV) \\ \hline
Gamma rays & - & - & - & 9.998$\times 10^{-1}$ & 8.346$\times 10^{-1}$ & 8.348 $\times10^{-1}$ \\ \hline
X rays & 6.662 & 1.356$\times 10^{-3}$ & 2.035$\times 10^{-4}$ & 6.670 & 1.360$\times 10^{-3}$ & 2.039$\times 10^{-4}$ \\ \hline
Annh. Photons & - & - & - & 1.120$\times 10^{-8}$ & 5.723$\times 10^{-9}$ & 5.11$\times 10^{-1}$ \\ \hline
Tot photons	& - & 0.0014 & - & - & 0.8360 & - \\ \hline
$\beta^{+}$	& - & - & - & 5.600$\times 10^{-9}$ & 1.019$\times 10^{-9}$ & 1.820$\times 10^{-1}$ \\ \hline
$\beta^{-}$ & - & - & - & 2.448$\times 10^{-4}$ & 2.03$\times 10^{-4}$ & 8.293$\times 10^{-1}$ \\ \hline
IC electrons & - & - & - & 2.448$\times 10^{-4}$ & 2.03$\times 10^{-4}$ & 8.293$\times 10^{-1}$ \\ \hline
Auger electrons & 5.415 & 3.989$\times 10^{-3}$ & 7.367$\times 10^{-4}$ & 5.422 & 4.001$\times 10^{-3}$ & 7.379$\times 10^{-4}$ \\ \hline
Tot electrons & - & 0.0040 & - & - & 0.0042 & - \\ \hline
Tot & - & 0.0053 & - & - & 0.8402 & - \\ \hline

\end{tabular}
\caption{Main decay data of the $^{53}$Mn and $^{54}$Mn radionuclides. For explanation see table \ref{tab:decay_1}.}
\label{tab:decay_2}
\end{center}
\end{table}

\subsection{Organ absorbed doses and effective dose due to $^{xx}$MnCl$_2$ injection}
Biodistribution data in healthy mice after $^{52g}$MnCl$_2$ injection \cite{hernandez} were used as input for dosimetric calculations. The percent of injected activity per gram of tissue ([\% IA/g]$_A$) in the main source organs (heart, liver, kidneys, muscle, pancreas and salivary gland) were used to evaluate the percent of injected activity into human organs ([\% IA/organ]$_H$) through the relative mass scaling method \cite{sparks}:

\begin{equation}
\left(\frac{\%IA}{organ}\right)_H = \left(\frac{\%IA}{g}\right)_A \left(\frac{OW_H}{TBW_H}\right) TBW_A\, 
\label{eq:2}
\end{equation}

where OW$_H$ is the human organ weight and TBW$_A$ and TBW$_H$ are the average total body weight for animal and human, respectively.                                 
For each source organ these data were plotted as a function of the post-injection (p.i.) time, from 1 to 13 days, and fitted to a tri-exponential equation, representing the phase of accumulation, retention and elimination of MnCl$_2$, with CoKiMo software \cite{melendez}. The number of disintegrations in each source organ due to $^{52g}$MnCl$_2$ injection was calculated by integration of the respective activity curve, considering the physical half-life of $^{52g}$Mn radioisotope. Similar calculations were performed for the other Mn radioisotopes  expected to be co-produced  by the investigated nuclear reactions.
Activity in the non-source organs, or remaining organs, was estimated by first assessing the total number of disintegrations in the body, on the base of  human biological half-life stated by Mahoney et al. \cite{mahoney}, and then subtracting to this number the number of disintegrations already ascribed to the source organs. This hypothesis is fundamental to properly assess the dose in the case of radioisotopes with a prolonged physical half-life, as already demonstrated in the case of $^{52}$MnCl$_2$ in a previous publication on this subject \cite{denardo2019}.
Absorbed dose calculations were performed with the OLINDA (Organ Level Internal Dose Assessment) software code version 2.2.0 \cite{stabin2012,stabin2005}, based on the RADAR method for internal dose estimation \cite{stabin2017} and the realistic NURBS-type male and female models \cite{stabin2012_2}, based on the standardized masses defined by ICRP 89 \cite{ICRP89}.  For each $^{xx}$Mn radioisotope, the effective dose coefficient, $ED^{^{xx}Mn}$ was obtained as the sum of the product of the organ equivalent dose per unit of administered activity, $D_{organ}^{^{xx}Mn}$, and the respective tissue-weighting factor, $w_{organ}$, recommended by ICRP 103 \cite{ICRP103},
\begin{equation}
ED^{^{xx}Mn}=\sum_{organs}{D_{organ}^{^{xx}Mn} \times w_{organ}} \, .
\label{eq:ED_xx}
\end{equation}
 
Finally, the total absorbed dose coefficient for different organs and the total effective dose coefficient ($ED_{tot}$) of MnCl$_2$, including the contribution of each $^{xx}$Mn radioisotope, were calculated at the different times after EOB, using the following equations:
\begin{equation}
D_{organ,tot}(t)= \sum_{xx} f_{^{xx}Mn}(t) \times D_{organ}^{^{xx}Mn}
\label{eq:abs_dose}
\end{equation}

\begin{equation}
ED_{tot}(t) = \sum_{xx} f_{^{xx}Mn}(t) \times ED^{^{xx}Mn}
\label{eq:effect_dose}
\end{equation}
where $f_{^{xx}Mn}(t)$ is the fraction of total activity corresponding to each radioisotope at the time $t$ after EOB, $A_{^{xx}Mn}(t)$. For easy reference, $f_{^{52g}Mn}(t)$ corresponds to the radionuclidic purity (RNP)
\begin{equation}
f_{^{52g}Mn}(t)=\frac{A_{^{52g}Mn}(t)}{\sum_{xx} A_{^{xx}Mn}(t)} .
\label{eq:rnp}
\end{equation}
The evaluation of the dose increase (DI) due to the presence of radiogenic impurities in the production route is introduced according to the following equation
\begin{equation}
DI(t)=\frac{ED_{tot}(t)}{ED^{^{52g}Mn}} \, ,
\end{equation}
and represents the ratio between the total effective dose ($i.e.$, including the impurities) and the effective dose due to an ideal injection of pure $^{52g}$Mn.

\section{Results}

\subsection{Cross sections}
\label{result_xs}

\begin{figure}[h!]
\begin{center}
\includegraphics{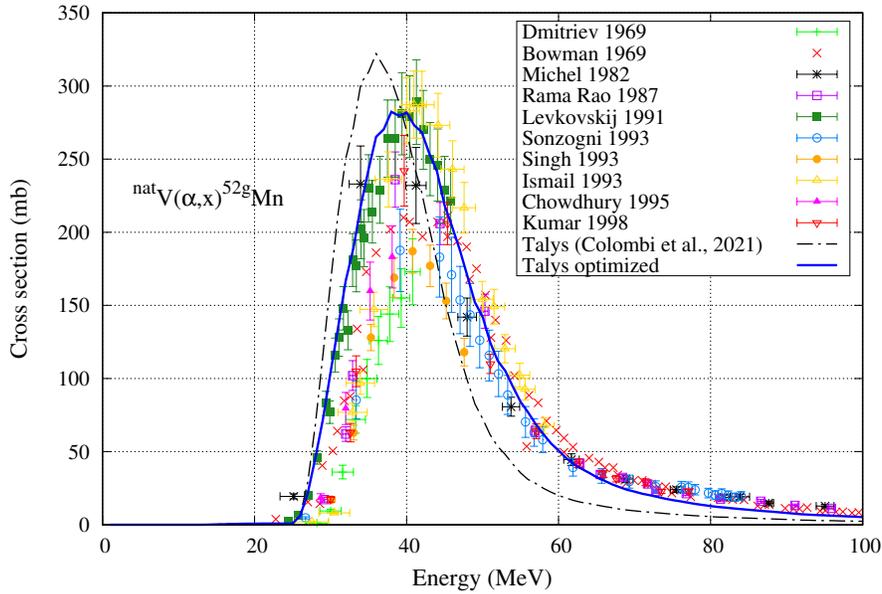}
\end{center}
\caption{$^{52g}$Mn cross section for the $^{nat}$V case, comparing the available experimental data and two theoretical curves: the newly TALYS optimized curve (solid line) and the one used by Colombi et al. (dash-dot line).}
\label{XS_52gMn}
\end{figure}

With the selection of models discussed in section \ref{sect:x-sect-Mns}, it was found necessary to 
optimize only the two level-density parameters with respect to the $^{52}$Mn compound, with the following values: $c=0.0$ MeV$^{-1/2}$ and $p=-1.0$ MeV. In practice, the $c=0.0$ MeV$^{-1/2}$ implies that no normalization factor has been introduced with respect to the original microscopic Hartree-Fock level density, while a shift of 1 MeV has been applied
towards higher energies, that could originate from shell-model or pairing effects. The resulting cross section is reported in figure \ref{XS_52gMn} by the solid line which represents a significant improvement with respect to the dashed line reporting the calculation adopted by Colombi et al \cite{colombi}. It must be added that, in the current result, the selection of the specific pre-equilibrium model with transition rates derived from the imaginary part of the optical potential was instrumental to correctly reproduce the cross-section at higher energies, in the region above 40 MeV. The standard models lead to a significant underestimation of the higher-energy data, as suggested also from the dash-dot line in figure \ref{XS_52gMn}. The available experimental data \cite{Dmitriev1969,Bowman1969,Michel1982,Rama1987,West1987,Sonzogni1993,Ismail1993,Singh1995,Chowdhury1995,Kumar1998,Peng1999} reported in the figure were taken from the EXFOR database \cite{EXFOR}.

Contaminant radioisotopes produced by this reaction route are characterized either by a very short, or by a very long, half-life. In particular, $^{48}$Mn, $^{49}$Mn, $^{50g/m}$Mn, $^{51}$Mn, and $^{52m}$Mn have half-lives shorter than one hour and their contamination, both in terms
of isotopic and radionuclidic purity, are negligible just after a few hours.
The stable $^{55}$Mn is produced in the electromagnetic channel with very low yield, and the quasi stable $^{53}$Mn (half-life $\sim$ 3.6×10$^{6}$ years) does not affect the radionuclidic purity in a significant way. Finally, we are left with the contaminant 
$^{54}$Mn, with an intermediate half-life of about 312 days. This could represent an obstacle limiting the radionuclidic purity of the production route.

\begin{figure}[h!]
\begin{center}
\includegraphics{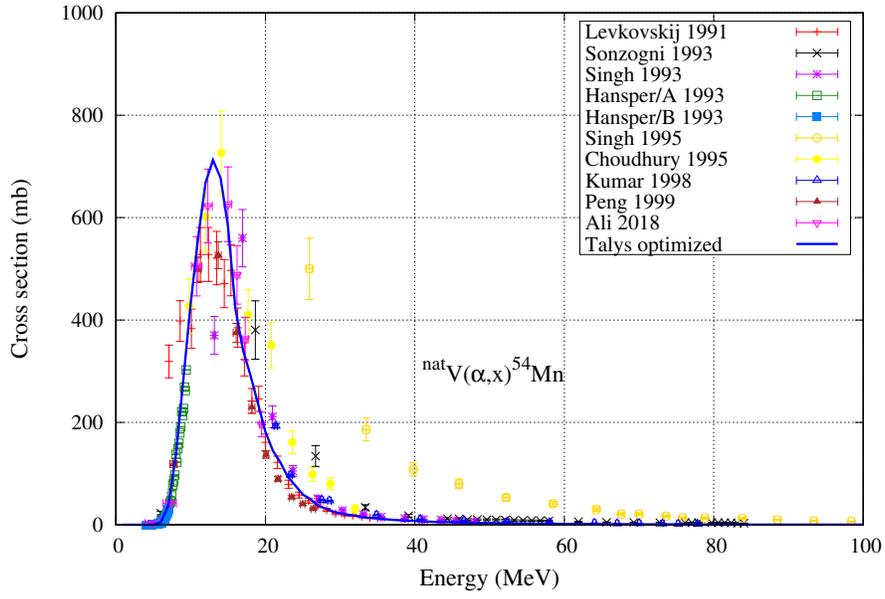}
\end{center}
\caption{$^{54}$Mn cross section  and theoretical curve obtained with the optimization used in this work.}
\label{XS_54Mn}
\end{figure}

The cross-section for $^{nat}$V($\alpha$,x)$^{54}$Mn is shown in figure \ref{XS_54Mn}, and the comparison between model and experimental data previously reported \cite{Sonzogni1993,Singh1995,Chowdhury1995,Kumar1998,Peng1999,Ali2018,Levkovskij1991,Singh1993,Hansper1993} is quite satisfactory. The calculations are rather stable with respect to the model variations in TALYS, while the experimental data are spread within a 30\% variability around the peak region. One must observe, however, that the relevant production region for $^{52g}$Mn is around 40 MeV, where the production of $^{54}$Mn is limited by the smallness of the corresponding cross section. Moreover, in this energy region, the data dispersion and the model variability are remarkably reduced.

\begin{figure}[!h]
\begin{center}
\includegraphics{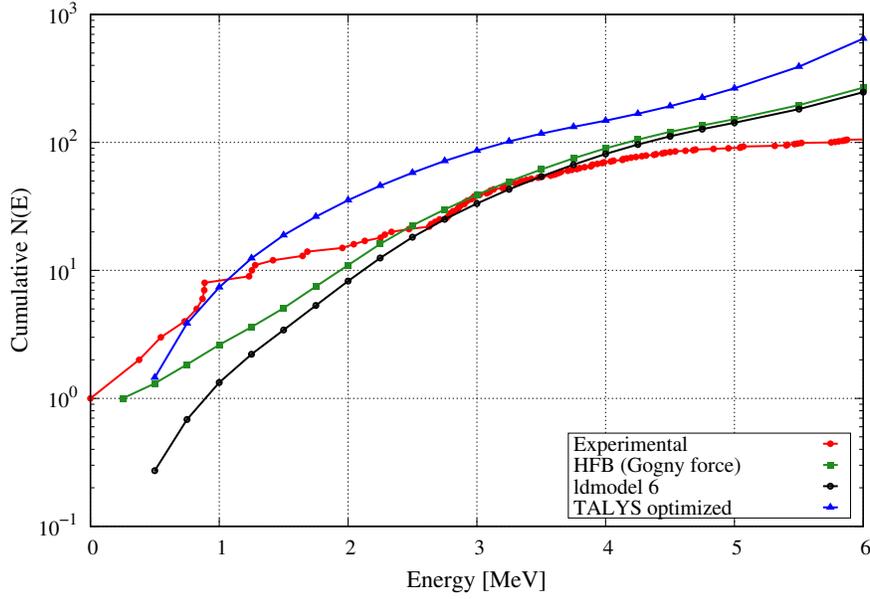}
\end{center}
\caption{Comparison of experimental and theoretical cumulatives of the number of levels for $^{52}$Mn.}
\label{fig:cumul}
\end{figure}

In figure~\ref{fig:cumul} the comparison between the experimental cumulative of the
number of levels and the theoretical ones is shown for  
the nuclide of interest. In particular, the green curve is the standard
theoretical HFB cumulative, with both parameters equal to zero. The  
black one is obtained by considering the values of $c$ and $p$ tabulated 
in TALYS, for the option \textit{ldmodel 6}, while the blue curve refers to the cumulative found in this work with the parameters associated to the optimized cross section. 
This solution seems in general to overestimate the trend of the cumulative and, in particular, in better agreement with data in the lower energy region. It is worth noting that the variation of the $c$ and $p$ parameters was not intended for a good description of nuclear level cumulatives. Rather, it has been carried out to optimize the agreement with the experimental cross section, since this is crucial for more accurate estimates of yields and purities in view of the specific nuclear applications described in this work.

\subsection{Yields}
\label{sect:yields}

The thick-target $^{52g}$Mn production yield has been calculated at the EOB instant, assuming 1 $h$ irradiation time and 1 $\mu A$ beam current. The calculations are developed following the method discussed previously by Canton et al \cite{canton-fontana}. 
The yield evaluation with $^{nat}$V target adopted the newly optimized TALYS calculation for the cross section, discussed in \ref{sect:x-sect-Mns}, which are more reliable than those derived by Colombi et al \cite{colombi}. Instead, for the $^{nat}$Cr target, the same cross sections calculated previously \cite{colombi} were used, since they turned out already reliable for the corresponding thick-target yield evaluation.

The two targets have been selected  both with a 200 $\mu$m thickness, corresponding for the $^{nat}$V target to the $\alpha$-beam energy window 48--33.9 MeV, while for $^{nat}$Cr to the 17--14 MeV range for the proton beam. The 200 $\mu$m thick target refers to a standard thickness suitable in both cases, and the corresponding energy ranges are selected to reduce the contamination. In table \ref{tab:act} the activity produced through the two routes are compared. The yields are similar, however the case $\alpha$-$^{nat}$V is significantly better, particularly for the smaller production of the long-lived $^{54}$Mn. The expected production of $^{51}$Mn is slightly larger, however this radionuclide is short lived, with an half-life of 46 min, and can be eliminated with a few hours of cooling.

\begin{table}[!htb]
\begin{center}
\begin{tabular}{|c|c|c|c|c|c|c|}
\hline
target&beam&Energy Range (MeV)&A($^{52g}$Mn)& A($^{51}$Mn) & A($^{53}$Mn) & A($^{54}$Mn) \\ 
\hline
$^{nat}$V& $\alpha$& [48 - 33.9]& 5.77 & 9.38 & 1.7$\times 10^{-8}$ & 3.0$\times 10^{-3}$   \\
\hline
$^{nat}$Cr& $p$& [17 - 14]& 4.40 & 8.82 & 7.5$\times 10^{-9}$ & 4.9$\times 10^{-3}$  \\
\hline
\end{tabular}
\caption{Comparison of the activities at EOB for the two production routes. The activities are given in  MBq/($\mu$A$\cdot h$).}
\label{tab:act}
\end{center}
\end{table}

\begin{figure}[h]
\begin{center}
\includegraphics{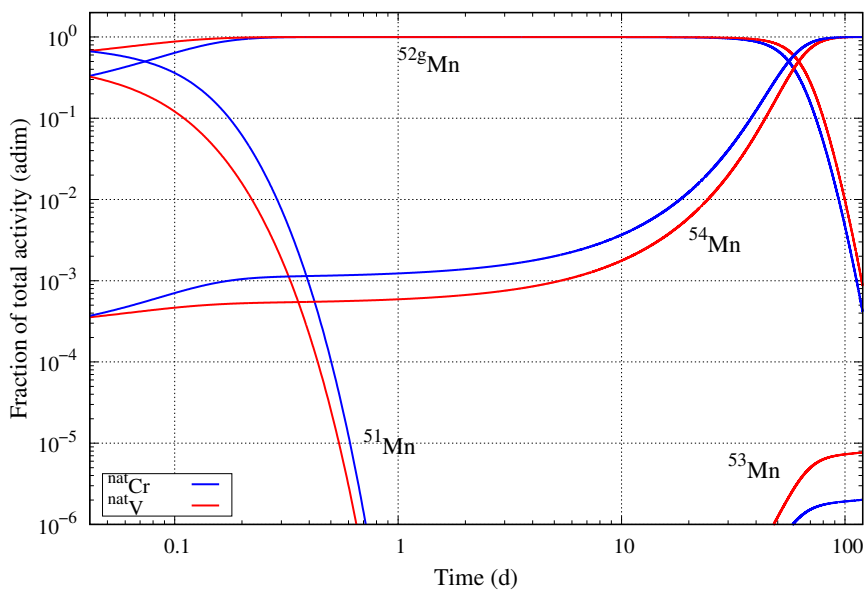}
\end{center}
\caption{Ratio of the activity of the given $^{xx}$Mn radionuclide with respect to the sum of all Mn ones. These fractions, in the case of $^{52g}$Mn, coincides with the definition of RNP.}
\label{fract_act}
\end{figure}

The time-evolution of the incidence of the radionuclides $^{52g}$Mn, $^{51}$Mn,  $^{53}$Mn, and $^{54}$Mn is reported in figure \ref{fract_act}. More specifically, the curves represent the fraction of total activity carried by each radionuclide, and for the $^{52g}$Mn case it corresponds to the standard definition of RNP (see (\ref{eq:rnp})). The figure compares the two routes of production and  clearly shows that the use of $\alpha$ on $^{nat}$V target provides yields with smaller contribution of the main contaminant, $^{54}$Mn.

\subsection{ Dosimetric calculations for $^{xx}$MnCl$_2$}
The organ activity curves obtained after fitting the percent of injected activity in the main source organs, [\% IA/organ]$_H$, calculated for male and female models through equation (\ref{eq:2}), evidenced that Mn$^{2+}$ shows a fast uptake for all the organs, followed by a slow wash-out, except for the salivary gland, where uptake remains quite stable. The number of disintegrations in the source organs per MBq of injected activity of $^{xx}$MnCl$_2$, obtained after integration of the organ activity curves, are reported in table \ref{tab:nuc_trans} for each radioisotope. The activity reported by Hernandez et al \cite{hernandez} as [\%IA/g] in “heart/blood” was assigned to “heart wall” and “heart contents” using equation (\ref{eq:2}) and the respective masses. As shown in table \ref{tab:nuc_trans}, for each radioisotope, the organ with the highest number of disintegrations is the liver, followed by the kidneys for the radionuclides with the shortest half-life ($^{51}$Mn and $^{52g}$Mn), by the salivary glands for the long-lived radionuclides $^{53}$Mn and $^{54}$Mn, due to the slow wash-out in this organ. Considering a total-body biological clearance based on the data of Mahoney and Small \cite{mahoney}, the calculated number of disintegrations in the remaining organs clearly increases with the physical half-life of the radioisotope. 

\begin{table}[!htb]
\begin{center}
\begin{tabular}{|c|c|c|c|c|c|c|c|c|}
\hline
&\multicolumn{2}{|c|}{[$^{51}$Mn]Cl$_2$} & \multicolumn{2}{|c|}{[$^{52g}$Mn]Cl$_2$} & \multicolumn{2}{|c|}{[$^{53}$Mn]Cl$_2$} & \multicolumn{2}{|c|}{[$^{54}$Mn]Cl$_2$} \\
\hline
Tissue&Female & Male & Female & Male & Female & Male & Female & Male \\
\hline
Heart contents & 1.37E-02 & 1.59E-02 & 0.882 & 1.00 & 3.63 & 4.12 & 3.43 & 3.89 \\ \hline
Heart wall & 9.38E-03 & 1.03E-02 & 0.596 & 0.647
& 2.46 & 2.66 & 2.32 & 2.51 \\ \hline
Kidneys & 2.64E-02 & 2.46E-02 & 2.24 & 2.07 & 4.75 &4.49 & 4.65 & 4.39 \\ \hline
Liver & 8.54E-02 & 9.02E-02 & 6.93 & 7.32 & 13.2 & 13.9 & 13.0 & 13.7 \\ \hline
Pancreas & 1.15E-02 & 1.10E-02 & 1.20 & 1.15 & 2.67 & 2.58 & 2.62 & 2.53 \\ \hline
Salivary gland & 2.71E-03 & 2.72E-03 & 0.601 & 0.600 & 6.43 & 7.64 & 5.52 & 6.36 \\ \hline
Remaining & 9.44E-01 & 9.38E-01 & 33 & 132 & 1100 & 1100 & 963 & 961.45 \\ \hline
\end{tabular}
\caption{Number of nuclear transitions (MBq-hr/MBq) in source organs per MBq of $^{xx}$MnCl$_2$, for female and male ICRP 89 phantoms.}
\label{tab:nuc_trans}
\end{center}
\end{table}

For unit activity of each radioisotope and for both male and female phantoms, data of table \ref{tab:nuc_trans} were used to calculate the absorbed doses to the organs and the ED based on ICRP 103 \cite{ICRP103} tissue weighting factors. Results are presented in table \ref{tab:abs_dose}. The organs receiving the highest doses are the kidneys, followed by the pancreas and the liver, in the case of $^{51}$MnCl$_2$, the pancreas, followed by the kidneys and liver, for $^{52g}$MnCl$_2$, both for male and female phantoms. The absorbed dose values obtained with the different radioisotopes are clearly correlated to their main decay data for dosimetric interest, reported in tables \ref{tab:decay_1} and \ref{tab:decay_2}.  Absorbed doses to different organs are 50-200 and 55-235 fold higher for $^{52g}$MnCl$_2$ compared to $^{51}$MnCl$_2$, respectively for male and female. This is due to the higher energy emitted by $^{52g}$Mn per nuclear transformation (nt), 3.5335 MeV/nt, compared to 1.9321 MeV/nt for $^{51}$Mn \cite{ICRP107}, and its longer half-life (see table \ref{tab:decay_1}). In the case of both $^{53}$MnCl$_2$ and $^{54}$MnCl$_2$, the organ receiving the highest dose becomes the salivary glands. For these long half-lives radioisotopes the absorbed dose to the other organs becomes almost uniform, due to the preponderant contribution of total body irradiation (see the number of counts in Remaining organs in table \ref{tab:nuc_trans}). Despite the long half-life of $^{53}$Mn, organ absorbed doses due to $^{53}$MnCl$_2$ are comparable to those due to $^{51}$MnCl$_2$, due to the very low radiation emission of this radioisotope, 0.0053 MeV/nt (ICRP107) (see table \ref{tab:decay_1}, \ref{tab:decay_2}). The total energy emitted per $^{54}$Mn decay is about 4 times lower compared to $^{52g}$Mn due to the absence of high-energy gamma emission, therefore organ absorbed doses due to $^{54}$MnCl$_2$ are only slightly higher than those due to $^{52g}$MnCl$_2$ despite the quite longer half life of $^{54}$Mn. As already reported for other radiopharmaceuticals \cite{stabin1997}, organ absorbed doses are in general about 20\% higher for female than for male, resulting in increased ED values between 33 and 38\%, depending on the radioisotope. As already reported in our previous publication \cite{denardo2019}, the ED values due to $^{51}$MnCl$_2$ injection are quite low and comparable to the ED of $^{18}$F-FDG (0.0192 mSv/MBq; gender-averaged value)\cite{stabin2017}, and about two orders of magnitude lower than those of $^{52g}$MnCl$_2$. The ED values due to $^{53}$MnCl$_2$ are less than a factor 3 higher than those due to $^{51}$MnCl$_2$, while those of $^{54}$MnCl$_2$ are about a factor 1.5 higher than those due to $^{52}$MnCl$_2$.
\\

\begin{table}[!htb]
\begin{center}
\footnotesize
\begin{tabular}{|c|c|c|c|c|c|c|c|c|}
\hline
&\multicolumn{2}{|c|}{[$^{51}$Mn]Cl$_2$} & \multicolumn{2}{|c|}{[$^{52g}$Mn]Cl$_2$} & \multicolumn{2}{|c|}{[$^{53}$Mn]Cl$_2$} & \multicolumn{2}{|c|}{[$^{54}$Mn]Cl$_2$} \\
\hline
Tissue&Female & Male & Female & Male & Female & Male & Female & Male \\
\hline
Adrenals & 1.83E-02 & 1.67E-02 & 2.65 & 2.27 & 4.22E-02 & 3.47E-02 & 3.39 & 2.69 \\ \hline
Brain & 1.14E-02 & 9.27E-03 & 1.43 & 1.15 & 4.22E-02 & 3.47E-02 & 2.46 & 1.98 \\ \hline
Breasts & 1.14E-02 & - & 1.35 & - & 4.22E-02 & - & 2.14 & -\\ \hline
Esophagus & 1.32E-02 & 1.09E-02 & 1.76 & 1.56 & 4.22E-02 & 3.47E-02 & 2.48 & 2.31 \\ \hline
Eyes & 1.14E-02 & 9.26E-03 & 1.43 & 1.14 & 4.22E-02 & 3.47E-02 & 2.45 & 1.95 \\ \hline
Gallbladder Wall & 1.49E-02 & 1.44E-02 & 2.41 & 2.24 & 4.22E-02 & 3.47E-02 & 3.24 & 2.75 \\ \hline
Left Colon & 1.37E-02 & 1.15E-02 & 2.22 & 1.90 & 4.22E-02 & 3.47E-02 & 3.57 & 2.97 \\ \hline
Small Intestine & 1.33E-02 & 1.13E-02 & 2.07 & 1.88 & 4.22E-02 & 3.47E-02 & 3.27 & 3.06 \\ \hline
Stomach Wall & 1.39E-02 & 1.19E-02 & 2.21 & 1.87 & 4.22E-02 & 3.47E-02 & 3.37 & 2.70 \\ \hline
Right Colon & 1.38E-02 & 1.14E-02 & 2.26 & 1.89 & 4.22E-02 & 3.47E-02 & 3.61 & 2.91 \\ \hline
Rectum & 1.30E-02 & 1.06E-02 & 2.12 & 1.74 & 4.22E-02 & 3.47E-02 & 3.66 & 2.99 \\ \hline
Heart Wall & 3.56E-02 & 3.02E-02 & 2.16 & 1.82 & 3.39E-02 & 2.78E-02 & 2.98 & 2.33 \\ \hline
Kidneys & 5.66E-02 & 4.71E-02 & 3.12 & 2.58 & 3.96E-02 & 3.33E-02 & 3.19 & 2.61 \\ \hline
Liver & 4.05E-02 & 3.34E-02 & 2.90 & 2.37 & 2.17E-02 & 1.78E-02 & 2.77 & 2.18 \\ \hline
Lungs & 1.33E-02 & 1.07E-02 & 1.82 & 1.44 & 4.22E-02 & 3.47E-02 & 2.83 & 2.20 \\ \hline
Ovaries & 1.31E-02 & - & 2.17 & - & 4.22E-02 & - & 3.71 & - \\ \hline
Pancreas & 5.75E-02 & 4.68E-02 & 3.70 & 3.00 & 5.12E-02 & 4.24E-02 & 3.78 & 3.11 \\ \hline
Prostate & - & 1.06E-02 & - & 1.72 & - & 3.47E-02 & - & 2.94 \\ \hline
Salivary Glands & 2.17E-02 & 1.83E-02 & 2.27 & 2.01 & 2.11E-01 & 2.07E-01 & 3.93 & 3.73 \\ \hline
Red Marrow & 1.01E-02 & 8.32E-03 & 1.84 & 1.47 & 4.22E-02 & 3.47E-02 & 3.05 & 2.46 \\ \hline
Osteogenic Cells & 8.01E-03 & 7.59E-03 & 1.88 & 1.55 & 1.54E-02 & 1.65E-02 & 3.25 & 2.81 \\ \hline
Spleen & 1.42E-02 & 1.11E-02 & 2.18 & 1.71 & 4.22E-02 & 3.47E-02 & 3.29 & 2.66 \\ \hline
Testes & - & 9.64E-03 & - & 1.33 & - & 3.47E-02 & - & 2.29 \\ \hline
Thymus & 1.31E-02 & 1.09E-02 & 1.87 & 1.49 & 4.22E-02 & 3.47E-02 & 2.98 & 2.32 \\ \hline
Thyroid & 1.19E-02 & 1.00E-02 & 1.60 & 1.45 & 4.22E-02 & 3.47E-02 & 2.68 & 2.45 \\ \hline
Urinary Bladder Wall & 1.19E-02 & 1.05E-02 & 1.56 & 1.70 & 4.22E-02 & 3.47E-02 & 2.65 & 2.94 \\ \hline
Uterus & 1.31E-02 & - & 2.14 & - & 4.22E-02 & - & 3.68 & - \\ \hline
Total Body & 1.35E-02 & 1.10E-02 & 1.70 & 1.37 & 4.34E-02 & 3.58E-02 & 2.72 & 2.28 \\ \hline
ED (ICRP 103) & 1.36E-02 & 1.02E-02 & 1.79 & 1.35 & 3.89E-02 & 2.82E-02 & 2.73 & 2.02 \\ \hline
\end{tabular}
\caption{The absorbed doses (mSv/MBq) calculated for $^{xx}$MnCl2 with the OLINDA v2.2 software for female and male ICRP 89 phantoms using the data of table \ref{tab:nuc_trans}, and ED  (mSv/MBq) values based on the ICRP 103 tissue weighting factors. }
\label{tab:abs_dose}
\end{center}
\end{table}

\section{Discussion}
In this work two production routes for $^{52g}$Mn have been compared, the standard low-energy $^{nat}$Cr($p$,x) and the alternative intermediate-energy $^{nat}$V($\alpha$,x) one, disregarded in the past reviews, and proposed only very recently. The former route has very consistent cross section data which are well reproduced by standard nuclear reaction codes such as TALYS, FLUKA, and EMPIRE, while the latter has measured data quite scattered and the standard nuclear reaction calculations do not perform so well.  
To improve the comparison, an optimized calculation within the TALYS code has been developed where the parameters of the nuclear level densities in the microscopic Hartree-Fock formalism have been adjusted to the measured cross sections. In addition, the exciton-based pre-equilibrium model that better described the higher energy tail of the cross section has been selected. Considering this new optimized cross-section, the comparative study between the two production routes, already performed by Colombi et al. \cite{colombi}, has been repeated and extended also to evaluate the effective doses imparted to simulated phantoms.
To ensure the safety of a radiopharmaceutical, the production methods should generate a minimal amount of radiogenic contaminants. The European Pharmacopoeia requires a radionuclidic purity limit of 99\% \cite{EANM2019}. In general, the time range that satisfies this requirement is reported in Tab.\ref{tab:rnp_di} for the two production routes. The lower time limit is due to the $^{51}$Mn fast decay, the upper one to the slow decay of $^{52g}$Mn. The comparison shows a longer time interval that matches the RNP limit for the $^{nat}$V case, and this is due to a larger production of $^{52g}$Mn and a lower $^{54}$Mn contamination, as can be inspected also from figure \ref{fract_act}. The large production of the short-lived $^{51}$Mn affects the purity only in the short term, while the $^{53}$Mn has a negligible impact well beyond the time interval of interest. On the other hand $^{51}$Mn could be considered an additional (fast-decaying) radionuclide for PET diagnostics \cite{brandt} to be used in co-prodution with $^{52g}$Mn. 

\begin{figure}[!h]
\begin{center}
\includegraphics[width=12cm]{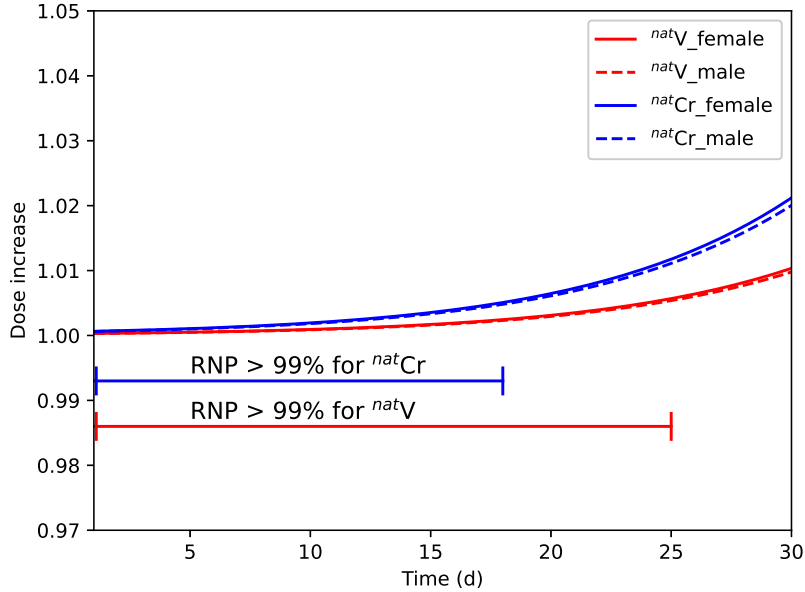}
\end{center}
\caption{Time evolution of the DI for the two production routes and for female/male phantoms. The blue and red lines refer to the $^{nat}$Cr($p$,x) and $^{nat}$V($\alpha$,x) reactions, respectively. The horizontal lines indicate the time ranges allowed by the 99\% RNP limits.}
\label{DI_tot}
\end{figure}

It is not sufficient to verify the fraction of radionuclidic impurities, but it is essential to quantify also the dose increase due to the presence of these impurities. Nowadays, there is not an established limit for the DI, but in general 10\% is considered a good starting value \cite{denardo2021,melendez2019}. The DI obtained for both production routes considering the anthropomorphic OLINDA phantoms are plotted in figure \ref{DI_tot} vs. the time between EOB and injection. In both cases the DI increases with time, however the use of $\alpha$ particles on $^{nat}$V targets guarantees a significantly longer time window where the total effective dose is less than 110\% with respect to an hypothetical pure $^{52g}$Mn injection. The time range satisfying this condition is reported in table \ref{tab:rnp_di}. The RNP limits, however, result in more stringent conditions than those obtained with the DI as also put in evidence in figure \ref{DI_tot}, showing the time range set from the RNP limits for both production routes and represented by the two horizontal bars. The DI obtained in the hypothesis of injecting $^{52g}$MnCl$_2$ within the 99\% time window is lower than 0.1\% and 0.3\%, respectively for $^{nat}$V and $^{nat}$Cr targets. It should however be remembered that the results concerning the DI are specific for each radiopharmaceutical, being dependent on its biokinetics.

\begin{table}[!htb]
\begin{center}
\begin{tabular}{|c|c|c|c|}
\hline
Target & Radionuclidic purity & \multicolumn{2}{|c|}{Dose Increase} \\ \hline
& Time range (day) & \multicolumn{2}{|c|}{Time range (day)} \\
& & female & male \\
\hline
$^{nat}$V & 0.28 - 24.35 & 0 - 50.21 & 0 - 50.78\\
\hline
$^{nat}$Cr & 0.29 - 18.28 & 0 - 44.14 & 0 - 44.71\\
\hline
\end{tabular}
\caption{Time range that satisfies the requirement of at least 99\% RNP for both production routes. Time range for which the DI is maintained within the limit of 10\%, for both production routes and both female and male OLINDA phantoms.}
\label{tab:rnp_di}
\end{center}
\end{table}

\section{Conclusions}

An improvement in the model reproduction of the $^{nat}$V($\alpha$,x)$^{52g}$Mn cross section, based on parameter optimization relevant for the nuclear level densities, allowed to 
make a more precise comparison between two production routes for the PET tracer $^{52g}$Mn.
The comparison between the $^{nat}$Cr(p,x) and the alternative intermediate-energy $^{nat}$V($\alpha$,x) production routes confirms that the use of natural vanadium targets leads to higher yield and higher RNP.
The comparison has been extended to the dosimetric impact due to the injection of the tracer $^{52g}$MnCl$_2$ to specific phantoms, taking into account the effect of the co-produced radionuclides $^{51}$Mn, $^{53}$Mn, and $^{54}$Mn. Assessments of the total doses imparted to the organs have been obtained, both in case of pure injection of one of the four radionuclides, as well as in case of mixtures expected from the modeling of the two production routes. Effective doses generated by the total doses over all organs, weighted by the appropriate tissue sensitivities, have been calculated and henceforth the dose increase, which describes the impact of the contaminants implied by the production route, has been derived. For both production routes, the DI is well within the 10\% limit for the entire range where the RNP is acceptable (i.e. greater than 99\%). Again, the $^{nat}$V($\alpha$,x)$^{52g}$Mn reaction provides a DI systematically lower than the one obtainable with $^{nat}$Cr($p$,x)$^{52g}$Mn, underlining the advantageous characteristics of the former reaction. 

Nevertheless, it must be acknowledged that the use of $\alpha$ particles at intermediate energies requires more complex (and less widespread) facilities than those providing low-energy proton beams, which can be provided by standard hospital cyclotrons.  Finally, the discussion has been limited here to the use of natural targets which are easily available and usually inexpensive. It is known that the use of enriched targets may improve the purity and yields, particularly in the Cr case, however this requires more expensive materials and additional complexity for target-recovery techniques.


{\bf Acknowledgments}

This research was funded within the INFN project CSN5-METRICS of the INFN Legnaro Laboratories.  
We thank Juan Esposito for stimulating scientific discussions.


\bibliographystyle{./medphy.bst}
\bibliography{manganese_medphy}

\end{document}